# Hole mobility increase in ultra-narrow Si channels under strong (110) surface confinement

Neophytos Neophytou* and Hans Kosina

Institute for Microelectronics, Technical University of Vienna, Austria

* e-mail: neophytou@iue.tuwien.ac.at

## Abstract

We report on the hole mobility of ultra-narrow [110] Si channels as a function of the confinement length scale. We employing atomistic bandstructure calculations and linearized Boltzmann transport approach. The phonon-limited mobility of holes in thin [110] channels can be improved by more than 3X as the thickness of the (110) confining surface is reduced down to 3nm. This behavior originates from confinement induced bandstructure changes that decrease the hole effective mass and the scattering rates. Our results provide explanations for recent mobility measurements in nanobelts of similar dimensions.

Silicon nanoscale channel devices have recently attracted significant attention as candidates for a variety of applications, such as high performance transistors, thermoelectric devices [1, 2], optoelectronic devices [3], and biosensors. At the nanoscale, enhanced confinement can alter the electronic dispersion and thus the electronic properties of a channel material, offering possibilities for performance optimization. At the nanoscale, however, enhanced confinement can alter the electronic dispersion and thus the electronic properties of a channel material. The transport and surface orientations and the confinement length scale are additional degrees of freedom in engineering device properties that could potentially improve mobility. This has been confirmed by measurements on a variety of nanoscale channel devices [4, 5, 6, 7].

Specifically for p-type Si channels, regarding the effect of orientation, it was shown that the (110)/[110] channels are largely advantageous compared to other surface/transport orientations [4, 5, 6]. Regarding the influence of confinement, theoretical works have indicated that the transport properties can in certain cases be largely improved as the confinement length scale in narrow ultra-thin-body (UTB) layers and nanowires (NWs) is reduced below ~10nm [8, 9, 10]. Recently, measurements in p-type [110] oriented nanobelt structures have also suggested that confinement can indeed improve carrier mobility [6, 7].

In this Letter we theoretically examine the effect of confinement on the hole low-field mobility of [110] Si ultra-thin 1D NWs, nanobelts, and 2D UTB layers. We consider confinement length scales from 3nm up to 13nm. We demonstrate that the hole low-field mobility in [110] channels may increase up to ~3X as the (110) confining side length is scaled down to 3nm. Such enhancement does not occur for (100) confinement. Our calculations are in good qualitative agreement with recent experimental data [6, 7]. We show that this behavior can be explained by confinement-induced bandstructure changes that largely reduce the effective mass of the subbands.

We employ the atomistic $sp^3d^5s^*$-spin-orbit-coupled (SO) tight-binding (TB) model [11] for electronic structure calculations and linearized Boltzmann transport theory. The TB model used is a compromise between computationally expensive ab-initio methods, and inexpensive but less accurate effective mass methods. Our calculations typically include up to 1500 atoms, a challenging, but achievable computational task within this model.

The electrical conductivity $\sigma$ follows from linearized Boltzmann theory as:

$$\sigma = q_0^2 \int_{E_V}^{\infty} dE \left(-\frac{\partial f_0}{\partial E}\right) \Xi(E), \qquad (1)$$

where the transport distribution function $\Xi(E)$ is defined as [12]:

$$\begin{aligned}\Xi(E) &= \sum_{k_x,n} v_n^2(k_x)\tau_n(k)\delta(E - E_n(k_x)) \\ &= \sum_n v_{k_x,n}^2(E)\tau_n(E)g_{1D}^n(E).\end{aligned} \qquad (2)$$

Here $v_{k_x,n}(E) = \frac{1}{\hbar}\frac{\partial E_n}{\partial k_x}$ is the bandstructure velocity, $\tau_n(k_x)$ is the momentum relaxation time for a state in the specific $k$-point and subband $n$, and $g_{1D}^n(E_n) = 1/2\pi\hbar|v_n(E)|$ is the density of states per spin for a 1D subband. The hole mobility $\mu_p$ is defined as $\mu_p = \frac{\sigma}{q_0 p}$, where $p$ is the hole concentration in the channel. The procedure is described in detail in [13].

We use Fermi's Golden rule to extract the momentum relaxation scattering rates. We include scattering due to elastic acoustic phonons (ADP), inelastic optical phonons (ODP), and surface roughness (SR), and use the full energy dependence for the momentum relaxation times. We assume an exponential autocorrelation function [14] for the roughness with amplitude $\Delta_{rms} = 0.2$nm and correlation length $L_C = 1.3$nm. For computational efficiency, some approximations were made: i) Confinement of phonons is neglected, and bulk dispersionless phonons are assumed. Instead, enhanced deformation potential values $D_{ODP}^{hole} = 10.5 \times 10^{10}$ eV/m, $D_{ADP}^{hole} = 5$ eV are employed, as is common practice for nanostructures [9, 15, 16]. This can influence the results by 10-20% [9, 17]. ii) Surface relaxation is neglected. iii) For SRS we derive the transition rate from the shift in the band edges $\Delta E_v / \Delta L$ with confinement [13, 14, 18]. As discussed by Uchida *et* al. [18], this is the strongest contribution to SRS in channels of a few nanometers in thickness. These commonly employed approximations. Although in certain cases they might be quite strong, it is believed that they affect the results only quantitatively. Qualitatively, our results depend mostly on the electronic structure and how it varies due to the change in geometry, which is the main focus of this work. As we show below, experimental mobility trends can be explained from these variations in electronic structure alone.

Trivedi *et* al. have measured the mobility of [110] transport oriented nanobelts under strong (100) surface confinement [7]. The nanobelt top-to-bottom surface orientation is (100), whereas the side surface orientation is (110). The corresponding *height* of the nanobelts is in [100] direction and was $H_{[100]}$=4.3nm, whereas the *width* is in [110] direction and was reduced gradually from $W_{[110]}$=169nm to 3.4nm. In Fig. 1 we show the mobility reported for three of these channels (square-red points). The solid line shows our simulation results for the phonon-limited mobility of nanobelts of the same height $H_{[100]}$=4.3nm, but with the width varying from $W_{[110]}$=13nm down to 3nm. The points labeled "A", show the calculated phonon-limited mobility for the 2D (100)/[110] UTB channel of thickness 4.3nm, which relates more to the nanobelt with $H_{[100]}$=4.3nm, $W_{[110]}$=169nm. Two simulation results are shown corresponding to two different sets of deformation potential parameters. For "P1" we use the parameters specified above, whereas for "P2" we use bulk parameters [19]. The result for "P2" is closer to the measured mobility of the wider nanobelt, which indicates that bulk deformation potential values might still be valid for thin-layers, as also mentioned by Donetti *et al.* [17]. What is important, however, is that regardless of the value used for the deformation potentials, qualitatively both measurements and simulations indicate a clear mobility increase <3X as the nanobelt is scaled down to an ultra thin nanowire. Even when including SRS in the calculation (dashed line) the qualitative behavior is still retained.

This behavior originates from a large bandstructure change under (110) confinement in p-type channels, which largely decreases the effective mass of the subbands in the [110] transport direction. To verify that the mobility improvement originates from the (110) confinement and is not a result of generic 2D to 1D confinement, Point B in Fig. 1 shows the calculated mobility of the (110)/[110] UTB channel with (110)-thickness of 4.3nm. This is a 2D structure, confined only by the (110) surface. The mobility reaches a value of $\mu$=1130 cm$^2$/Vs. Comparing Point B to Point A [the (100)/[110] UTB channel], it is clear that for [110] oriented p-type channels, strong (110) confinement is beneficial to the mobility, whereas (100) confinement is not.

Figure 2 shows how this bandstructure modification takes place under strong (110) confinement. Figure 2a shows the dispersion relation of a square NW with cross section 3nm x 3nm, in the [110] transport orientation with the band-edge shifted to $E$=0eV. Figure 2b shows the highest subband ("envelope" of the dispersion) of NWs with geometrical features $W_{[110]}$=12nm, $H_{[100]}$=3nm [circled-red labeled (12,3)], $W_{[110]}$=3nm, $H_{[100]}$=12nm [squared-blue labeled (3,12)], and $W_{[110]}$=12nm, $H_{[100]}$=12nm [triangle-black labeled (12,12)]. The (12,3) NW is under stronger (100) confinement, the (3,12) NW under stronger (110) confinement, and the (12,12) NW under weak confinement from both surfaces. For reference we show the highest subband of the 3nm x 3nm NW (black), which is under strong confinement from both surfaces. The 3nm x 3nm NW has a dispersion with large curvature (small effective mass of $m^*$~0.2$m_0$), which provides large carrier velocities and improved mobilities. In Fig. 2b, the dispersion of the (12,12) NW has a smaller curvature (larger $m^*$~0.8$m_0$), and very similar to the curvature of the (12, 3) NW. On the other hand, the (3,12) NW has a curvature very similar to that of the (3,3) NW. Clearly the (110) confinement is the one responsible for providing the large dispersion curvatures, whereas the (100) confinement does not change the curvature of the dispersion significantly from the (12,12) NW.

This analysis holds for 2D UTB channels as well. Figure 3 shows the calculated low-field phonon-limited mobility for holes in 2D UTB layers under (100) and (110) surface confinement as a function of the layer thickness, T. For both confinements we show both [100] and [110] transport orientations. We vary the UTB layer thickness from T=10nm down to 3nm. For the larger layer thicknesses, the (110)/[110] channel is highly advantageous compared to all other channels, with almost ~2X higher mobility. More importantly, however, as the thickness is reduced to 3nm, the mobility of the (110)/[110] channel increases by ~3X, an improvement not observed for the rest of the UTB channels. This is in qualitative agreement with measurements in MOSFET devices [4, 5]. The mobility of the (110)/[100] channel also increases, but the increase is smaller. Transport in [110] direction benefits the most from the effective mass reduction under (110) confinement, whereas in [100] lesser. On the other hand, the (100) surface channels not only provide the lowest mobility, but a small degradation is also observed as the thickness is reduced. In contrast to (110) confinement, (100) confinement does not offer any advantage. We note here that the carrier mobility and carrier velocity presented in Ref. [8], although correlated, do not show the same quantitative behavior. The variations in mobility with confinement and orientation, especially for the (110)/[110] channel are much larger than the variations in velocity.

The dashed line in Fig. 1 shows the mobility calculation for which phonon scattering *and* SRS are included. The detrimental effect of SRS on the mobility of the nanobelts originates from the roughness along the surface with the largest area, i.e. from the (100) surface. Figure 4 examines the qualitative influence of the (100) and (110) surrounding surfaces on SRS by showing the ratio of the phonon- plus SRS-limited mobility to the phonon-limited mobility for two nanobelt categories. The first is the (100)/[110] nanobelt as shown in Fig. 1, and the second one is the (110)/[110] nanobelt, for which the width in [110] direction remains constant at $W_{[110]}$=4.3nm, whereas the height in [100] direction changes from $H_{[100]}$=13nm down to $H_{[100]}$=3nm. These are orthogonal situations with respect to confinement.

Results are shown for two different $\Delta_{rms}$ values, 0.2nm and 0.48nm. Interestingly, the nanobelts with (110) confinement (circle-red lines) are less affected by SRS compared to the (100) surface confined channels (triangle-blue lines). This can be explained from the fact that (110) confinement causes a smaller band edge shift $\Delta E_V / \Delta W$ compared to (100) confinement, as shown in the inset of Fig. 4 (right-most region). For (110) nanobelts, as the height in [100] direction is reduced, the band edges rapidly shift, whereas for (100) nanobelts, as the width in [110] direction is reduced, the band edges do not shift any further. The fact that (110) confinement does not affect the band edges strongly indicates a larger (110) confinement effective mass. Overall, not only (110) confinement improves mobility, but also offers stronger immunity to SRS.

A second important observation is that as the width of the nanobelt is reduced and the structure geometry shifts from 2D thin-layers towards 1D NWs, the effect of SRS is gradually reduced. This suggests that SRS is less effective for NWs than for 2D UTB layers as also pointed out in Ref. [20]. The reason is that in wider nanobelts a larger amount of modes participate in transport, and SRS scattering is enhanced through coupling of these modes.

In summary, we investigated the effect of (110) surface confinement on the low-field mobility of [110] p-type Si nano-channels (1D NWs, nanobelts, and 2D UTB layers) using atomistic electronic structures and Boltzmann transport. We find that the (110)/[110] channel shows a significant performance advantage, and that the phonon-limited mobility in such channel undergoes a large increase as the (110) confinement increases. This can be explained by a reduction of the effective mass with confinement, which is possible under (110), but not under (100) confinement. Furthermore, the (110) surface shows stronger immunity to the detrimental effect of SRS, which overall makes the (110)/[110] channel ideal for ultra-thin p-type channels. Our results provide explanations for recent experimental mobility measurements in p-type nanobelt and UTB channels in which mobility enhancement under (110) confinement was reported [4, 5, 6, 7].

*Acknowledgement:* This work was supported by the Austrian Climate and Energy Fund, contract No. 825467.

# References

* e-mail: neophytou@iue.tuwien.ac.at

Figure 1:

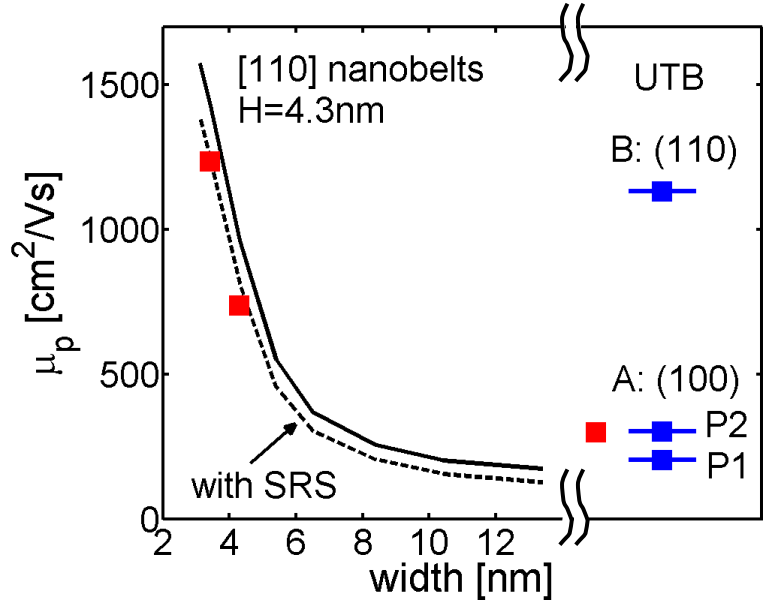

Figure 1 caption:

Hole low-field mobility for nanobelts of height $H_{[100]}$=4.3nm vs. their width $W_{[110]}$. Solid line: Phonon-limited. Dashed-line: Phonon scattering plus SRS. Square-red points: Measurements from Ref. [7]. Point A: Phonon-limited mobility of (100)/[110] UTB layer with thickness $T_{[100]}$=4.3nm. P1: $D^{holes}_{ODP} = 10.5 \times 10^{10} \text{eV}/\text{m}$, $D^{holes}_{ADP} = 5 \text{ eV}$. P2: $D^{holes}_{ODP} = 6 \times 10^{10} \text{eV}/\text{m}$, $D^{holes}_{ADP} = 5 \text{ eV}$. Point B: Phonon-limited mobility of (110)/[110] UTB layer with thickness $T_{[110]}$=4.3nm.

Figure 2:

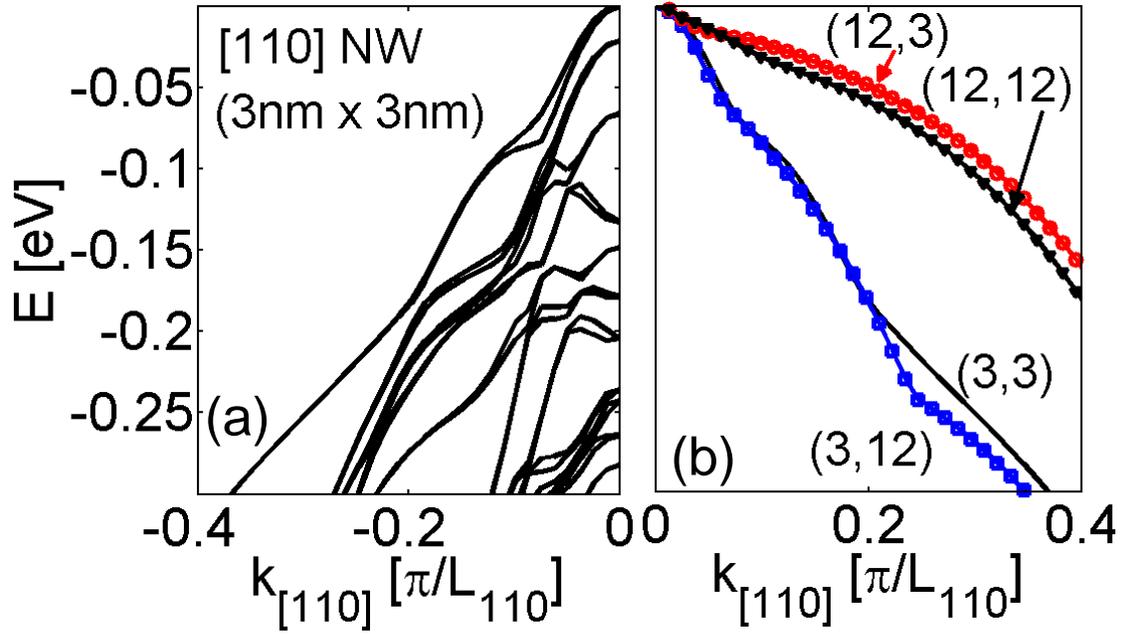

Figure 2 caption:

Bandstructure variation with confinement for p-type [110] NWs. (a) Hole dispersion relation for the 3nm x 3nm NW. (b) Highest dispersion subband (envelope) for NWs with dimensions labeled ($W_{[110]}$, $H_{[100]}$).

Figure 3:

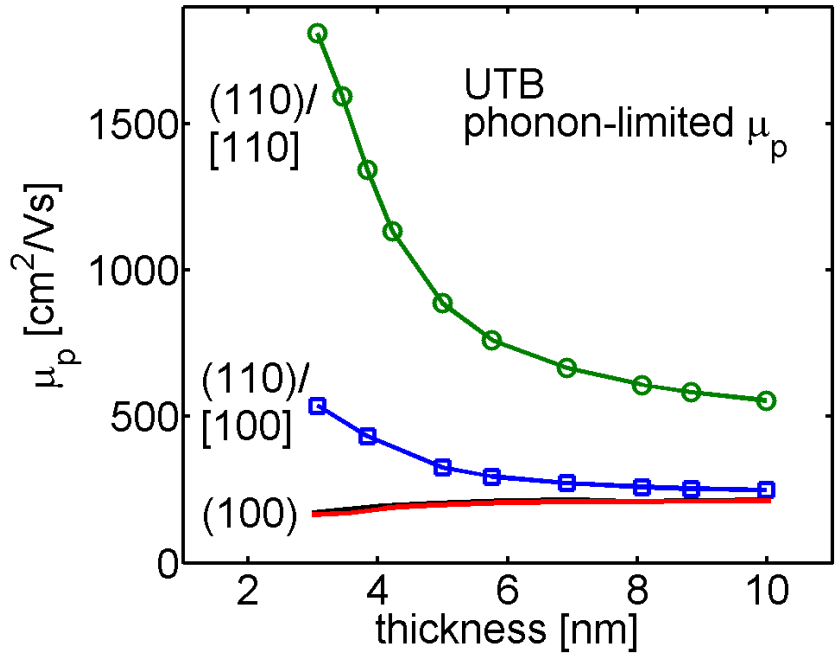

Figure 3 caption:

Low-field phonon-limited mobility of holes for UTB layers vs. their thickness. Results for (100) and (110) surfaces, and both [100] and [110] transport orientations for each surface are shown.

Figure 4:

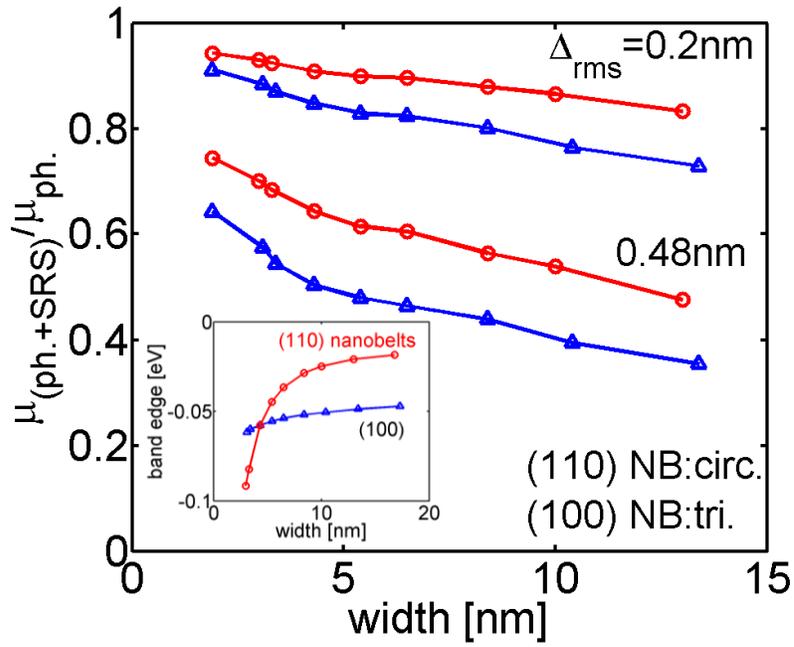

Figure 4 caption:

The ratio of the phonon- plus SRS-limited mobility to the phonon-limited mobility for the (100) and (110) confined p-type [110] transport nanobelts (NB). Inset: The band edges of the two nanobelt categories with thickness variation.